\documentclass[sigconf]{acmart}
\usepackage{booktabs}
\usepackage{array,tabularx}
\newcolumntype{Y}{>{\raggedright\arraybackslash}X}
\usepackage{enumitem}
\usepackage{xcolor} 
\usepackage{colortbl}
\usepackage{makecell} 
\usepackage{tikz}
\usetikzlibrary{arrows.meta,positioning,calc,shapes.symbols}
\usepackage[T1]{fontenc}

\usepackage{array}
\newcolumntype{P}[1]{>{\centering\arraybackslash}p{#1}}

\AtBeginDocument{%
  }

\usepackage{todonotes}
\newcounter{todocounter}

\definecolor{HeatBase}{RGB}{47,85,151} 

\setcopyright{acmlicensed}
\copyrightyear{2018}
\acmYear{2018}
\acmDOI{XXXXXXX.XXXXXXX}
\acmConference[Conference acronym 'XX]{Make sure to enter the correct
  conference title from your rights confirmation email}{June 03--05,
  2018}{Woodstock, NY}
\acmISBN{978-1-4503-XXXX-X/2018/06}

\begin{document}

\newcommand\mynote[1]{\marginpar[\scriptsize \textcolor{red}{\hfill\bf #1}]{\scriptsize \textcolor{red}{\bf #1}}}
\newcommand{\myitem}{\vspace*{-2.5mm}\item}
\newcommand{\mysection}{\vspace*{-1.5mm}\section}
\newcommand{\mysubsection}{\vspace*{-1mm}\subsection}
\newcommand{\mysubsubsection}{\vspace*{0mm}\subsubsection}
\newcommand{\myparagraph}{\vspace*{-2mm}\paragraph}
\newcommand{\mybibitem}{\vspace*{0mm}\bibitem}

\newcommand\noteIZ[1]{\marginpar[\scriptsize \textcolor{red}{\hfill\bf #1}]{\scriptsize \textcolor{red}{\bf #1}}}
\newcommand\noteMZ[1]{\marginpar[\scriptsize \textcolor{violet}{\hfill\bf #1}]{\scriptsize \textcolor{violet}{\bf #1}}}
\newcommand\trackWW[1]{\textcolor{red}{#1}}
\newcommand\trackDH[1]{\textcolor{blue}{#1}}
\newcommand\trackMZ[1]{\textcolor{violet}{#1}}
\newcommand\trackIF[1]{\textcolor{violet}{#1}}

\newcommand{\eg}{e.g., }
\newcommand{\ie}{i.e., }
\title{Invisible Load: Uncovering the Challenges of Neurodivergent Women in Software Engineering}


\author{Munazza Zaib}
\affiliation{%
  \institution{Monash University}
  \city{Melbourne}
  \country{Australia}}
\email{munazza.zaib@monash.edu}

\author{Wei Wang}
\affiliation{%
  \institution{Monash University}
  \city{Melbourne}
  \country{Australia}}
\email{wei.wang7@monash.edu}

\author{Dulaji Hidellaarachchi}
\affiliation{%
  \institution{RMIT}
  \city{Melbourne}
  \country{Australia}}
\email{dulaji.hidellaarachchi@rmit.edu.au}

\author{Isma Farah Siddiqui}
\affiliation{%
  \institution{Monash University}
  \city{Melbourne}
  \country{Australia}}
\email{ismafarah.siddiqui@monash.edu}

\renewcommand{\shortauthors}{Trovato et al.}

\begin{abstract}

Neurodivergent women in Software Engineering (SE) encounter distinctive challenges at the intersection of gender bias and neurological differences. To the best of our knowledge, no prior work in SE research has systematically examined this group, despite increasing recognition of neurodiversity in the workplace. Underdiagnosis, masking, and male-centric workplace cultures continue to exacerbate barriers that contribute to stress, burnout, and attrition. In response, we propose a hybrid methodological approach that integrates InclusiveMag’s inclusivity framework with the GenderMag walkthrough process, tailored to the context of neurodivergent women in SE. The overarching design unfolds across three stages, scoping through literature review, deriving personas and analytic processes, and applying the method in collaborative workshops. We present a targeted literature review that synthesize challenges into cognitive, social, organizational, structural and career progression challenges neurodivergent women face in SE, including how under/late diagnosis and masking intensify exclusion. 
These findings lay the groundwork for subsequent stages that will develop and apply inclusive analytic methods to support actionable change.
\end{abstract}



\maketitle

\section{Introduction}
The landscape of the technology sector is evolving rapidly as it is becoming more inclusive and diverse \cite{rodriguez2021perceived,austin2017neurodiversity}. As workplace inclusion efforts expand, neurodiversity is encompassing natural variations in cognition and perception and is gaining recognition for the unique strengths neurodivergent individuals bring to problem-solving and innovation in technical fields \cite{NeurodiversityHennekamEtAl2024, hoogman2020creativity, austin2017neurodiversity, Grandin2022}. The concept of neurodiversity has expanded beyond individuals with formal diagnoses (e.g., autism, ADHD, learning disorders) to include a broader population of those who self-identify as neurodivergent \cite{dwyer2022neurodiversity}. In Australia, an estimated 30–40\% of the population identifies as neurodivergent \cite{AUsupport}, and this trend is reflected in software engineering (SE) domain, with the Stack Overflow Developer Survey (2018–2022) reporting 4.27\% of respondents identifying with autism/ASD and 10.27\% with concentration or memory disorders (e.g., ADHD) among over 71,000 participants \cite{Stack2022}.

Employers and researchers, particularly within SE, are beginning to recognise neurodivergent individuals as valuable contributors, offering distinct strengths that can enhance team performance and innovation \cite{NDInAgileTeamsGamaEtAl2023, austin2017neurodiversity}. Major tech companies SAP, Hewlett Packard Enterprise, and Microsoft have initiated neurodivergent hiring programs to leverage these strengths \cite{austin2017neurodiversity}. However, the benefits of these initiatives are not equally experienced by all. Women with neurodiverse conditions often face compounded challenges, navigating both gender bias and the stigma associated with neurological difference \cite{CamouflagingLaiEtAl2017, cook2021camouflaging}. Conditions such as autism and ADHD have historically been underdiagnosed in women, leading to a lack of recognition, delayed diagnosis, and minimal support during key developmental and career stages \cite{QualitativeStudyMilnerEtAl2019,LateDiagnosisBargielaEtAl2016}. These burdens are further intensified by masking pressures. Many neurodiverse women report \textit{“camouflaging”} their traits to meet expectations of both femininity and professional competence \cite{CamouflagingLaiEtAl2017, cook2021camouflaging,hull2020female}. In male-dominated SE workplaces, where extroverted and fast-paced communication styles are often valorized, such continuous self-monitoring can be exhausting \cite{hennekam2024neurodiversity}.  Over time, this constant self-monitoring contributes to mental health decline, burnout, and diminished workplace participation~\cite{jahandideh2025low, mantzalas2022autistic, raymaker2020having}. 

While recent SE research has begun exploring inclusivity from a neurodiversity perspective \cite{rodriguez2021perceived, marquez2024inclusion}, the \textit{intersection of gender and neurodiversity remains significantly underexplored}. Most studies have focused either on gender diversity or general developer well-being, but not on the unique lived experiences of \textbf{\textit{neurodivergent women in SE}} \cite{rodriguez2021perceived, gama2023understanding}. This paper is part of an effort to reimagine inclusion in software engineering, not through isolated accommodations, but through reengineering the very processes that shape how teams work, evaluate and design. Our project asks: \textbf{\textit{How can software engineering practices be systematically redesigned to support the sustained participation, visibility, and well-being of neurodivergent women?}} To begin addressing this question, we focus on the Scope phase of our methodology, using a comprehensive literature review to surface the lived challenges neurodivergent women encounter across SE work.

This position paper argues that the barriers faced by ND women in SE expose deeper structural misalignments, where workplace norms, performance expectations, and design practices are still shaped around narrow definitions of communication, productivity, and competence. These challenges will not be solved by retrofitting accessibility onto existing systems. We advocate for a \textit{\textbf{fundamental shift toward systematic, intersectional inclusion}}, where design-for-diversity is embedded into the core of how teams hire, evaluate, and collaborate. To support this, we adopt and adapt the InclusiveMag meta-method \cite{mendez2019gendermag} and the GenderMag \cite{burnett2016gendermag} walkthrough framework as the foundation for a structured, phased approach. Establishing this foundation of documented challenges is essential for enabling future phases of research that will design and test inclusive analytic methods to support sustainable change in SE.


\section{Related Work}
 Kapp et al. \cite{kapp2013deficit} formalized neurodiversity as a \emph{deficit-difference} synthesis, recognizing that neuro-cognitive differences can involve disabling barriers while also representing legitimate forms of diversity. This further became popular in education and workplace contexts \cite{armstrong2010neurodiversity}, while Den Houting \cite{den2019neurodiversity} provided critical clarification of its meaning and misconceptions. In our paper, neurodiversity is considered as population-level differences such as autism and ADHD, where inclusion is achieved by adjusting environments and practices \cite{kapp2013deficit, den2019neurodiversity}.

\subsection{Neurodiversity in Women}
Historically, the neurodiversity spectrum, including ADHD and autism was framed through a male lens as it was thought to affect only boys \cite{IsueswithFemalescCiliaEtAl2023}. Because women and girls were largely absent from research, diagnostic criteria were developed based on male presentations of these conditions, which led to the under and late-diagnosis of women \cite{PresenttaionofAutismDriverEtAl2021, SexDifferencesPengEtAl2023}. As visibility has grown, more women are receiving diagnoses in adolescence or adulthood. However, this recognition tends to occur later and often after prolonged self compensation \cite{AustisticWomenChesterEtAl2019, LateDiagnosisBargielaEtAl2016}. A central mechanism in this delay is \emph{camouflaging}, the masking and compensation of autistic traits to meet social expectations \cite{hull2020female}. A study conducted by Lai et al.\cite{CamouflagingLaiEtAl2017} indicate that women on average score higher on camouflaging than men when it is quantified as a discrepancy between internal traits/social cognition and observable social behaviour, and that camouflaging correlates with depressive symptoms and sex-differentiated neural associations. A systematic review across 29 studies further indicates that camouflaging is prevalent, specially among verbally able adults and women, and robustly associated with anxiety and depression even though measures vary (self-report vs discrepancy indices) and longitudinal designs remain scarce \cite{cook2021camouflaging}. Complementing these findings, a quality study with late-diagnosed women explains that years of \emph{pretending to be normal}, gendered clinical blind spots and greater vulnerability during undiagnosed periods demonstrated concrete life-course costs of delayed recognition \cite{LateDiagnosisBargielaEtAl2016}. 

While increased awareness is shifting diagnostic trends \cite{AustisticWomenChesterEtAl2019}, girls and women remain disproportionately affected by delayed diagnoses across the neurodiversity spectrum and the downstream impacts are still under-examined \cite{NDWomenHoldenEtAl2025}. For many, limited support, the drain of having to mask to appear neurotypical and missed opportunities to design accommodating environments contribute to significant mental-health burdens \cite{CamouflagingLaiEtAl2017}. These issues intensify around major life transitions, from childhood to adulthood, relationships, and motherhood, where it is highlighted that autistic mothers report sensory overload and healthcare communication barriers and identify predictability and peer networks as protective \cite{pohl2020comparative}. This evidence suggests that process and environment fit, rather than motivation or capacity (their cognitive ability, intelligence, or competence), shape outcomes for ND women across the lifespan.

\subsection{Neurodiversity in SE Research}
Research on neurodiversity in SE has grown in recent years, highlighting both the strengths and barriers experienced by ND developers. An early large-scale Microsoft study found neurodiverse employees faced challenges with open-plan noise, frequent interruptions, and fast-paced communication, while also excelling in attention to detail, persistence, and pattern recognition \cite{Morris2015}. Later studies extended this perspective into Agile teams, when Gama and Lacerda \cite{NDInAgileTeamsGamaEtAl2023} found that stand-ups, rigid sprint ceremonies and unspoken collaboration norms often created exclusion while practices such as predictable cadence, artifact-first communication and opt-in pairing could make processes more accessible. 

Apart from these, task-specific studies have provided further granularity. Sasportes \cite{SasportesEtAl2024Challenges} found that autistic developers often excel in bug detection and detail checking but face barriers from implicit social norms, unstructured review protocols, and unclear tool design. Newman et al. \cite{newman2025get} focused on ADHD in professional programming and demonstrated that code review is particularly high-load with difficulties linked to abstraction, working memory and time management. They also documented compensatory strategies (e.g., chunking tasks, externalizing information, timers) that remain under-supported in organizational contexts. Analysis of the 2022 Stack Overflow developer survey by Verma et al. \cite{Liebel2025} indicated that neurodivergent developers compared to neurotypical peers reported more challenges with collaboration, coordination and interruptions suggesting systemic misfit in team-based practices. Menezes et al. \cite{da2025felt} highlighted how ND practitioners often felt pressured to \emph{"give 100\% all the time"}, linking this to burnout, fatigue and attrition risks. Costello et al. \cite{Costello2021Professional} in a review of autistic professionals' career pathways pointed to barriers in hiring, performance evaluation and workplace communication. Evidence from computing education further supports a fit-over-deficit interpretation. Graafsma et al. \cite{graafsma2022no} identified that autistic traits did not predict programming outcomes once cognitive skills were controlled, implying that differences in professional contexts are less about ability and more about environmental and organizational fit. Taken together, SE research shows that ND developers face barriers concentrated in communication load, interruptions, unstructured collaboration rituals and cognitively demanding tasks in SE \cite{Grischa2024Challenges}. These challenges map directly to the kinds of social and organizational pressures where ND women due to camouflaging, late diagnosis and gendered expectations are likely to experience intensified exclusion and burnout. Yet, current research are almost entirely \emph{gender-blind} as they rarely disaggregates by gender or examine how neurodiversity interacts with the gendered dynamics of male-dominated SE workplaces. This gap limits our ability to design interventions that address the compounded challenges faced by ND women in SE.

\section{Our Approach}

\begin{figure}[htb!]
\begin{center} 
\includegraphics[width=\linewidth]{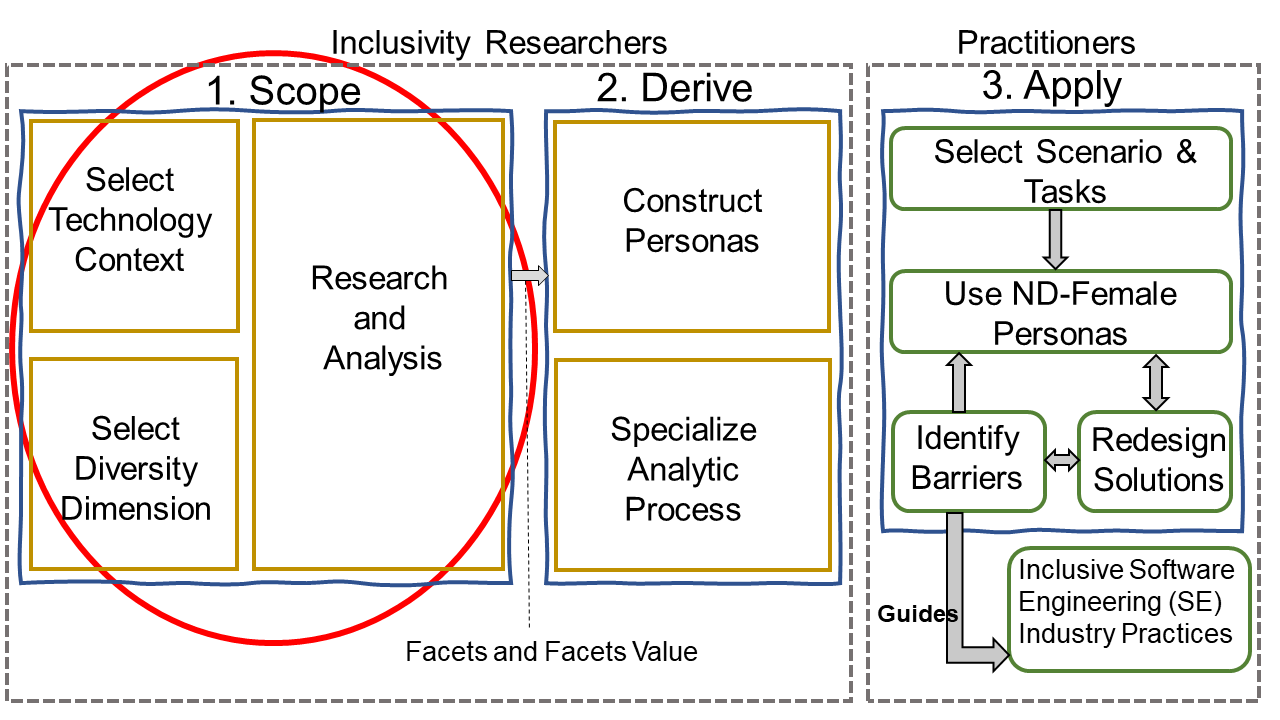}
\end{center}
\caption{Hybrid InclusiveMag–GenderMag Method: Supporting Neurodivergent Female Software Engineers, adapted from InclusiveMag \cite{mendez2019gendermag}}
\label{fig:architecture}
\vspace{-3mm}
\end{figure}

We are currently in the \textit{Scope phase} (as highlighted in Figure~\ref{fig:architecture}) of a three-stage methodology; \textbf{Scope, Derive, and Apply}, through which we apply and adapt the InclusiveMag meta-method to investigate the lived experiences of ND Women in SE (see Figure \ref{fig:architecture}). This phase focuses on defining the context and target population via a literature review and an exploratory survey. The insights gathered will inform the development of validated facets, personas, and task scenarios in the following stages. This work applies the InclusiveMag method, a meta-framework that enables inclusivity researchers to generate new inclusive design methods \cite{mendez2019gendermag}. InclusiveMag has been previously used in various domains, including age inclusivity \cite{mcintosh2021evaluating} and student inclusivity \cite{letaw2021changing,morreale2025we}. It was inductively derived by generalising the structure and principles of GenderMag \cite{burnett2016gendermag}, a method developed to uncover gender biases in software. We integrate InclusiveMag’s structured inclusivity framework with the GenderMag walkthrough process, tailoring it to better reflect the strengths, challenges, and needs of ND-Women in SE.

\textbf{\textit{Step One: Scope – Defining the Context and Target Population.}} We conducted a comprehensive literature review across SE, HCI, and neurodiversity studies to identify recurring barriers, strengths, and adaptation strategies for ND women in SE. This helped us to establish the five challenge categories used in the remainder of the paper. Based on the insights from our targeted literature review, the immediate next step is to conduct an exploratory survey study focusing on ND Women practitioners, to validate our findings and get a deeper understanding of these challenges. The survey responses will be used to refine and prioritize the challenges according to inclusion criteria \cite{hu2021toward}, ensuring that the framework is grounded in both empirical evidence. 

\textbf{\textit{Step Two: Derive – Personas and the Adapted Analytic Process.}} In phase 2, we will employ large language models (LLMs) as a scaffolding tool to generate draft personas that instantiate these challenges in situated scenarios. To mitigate risks of bias, stereotyping or reductive portrayals, we use them only as prompts rather than fixed labels, and each persona is grounded in evidence from our stage 1 review and survey results. The LLM outputs will be subjected to stereotype filtering, evidence anchoring and expert auditing to minimize bias and ensure fidelity. Based on the drafts, we will then refine and extend them by integrating the structured facets identified through the survey. This process yields a small, diverse set of personas that capture varied profiles of ND-women in SE while explicitly including cautions against essentialism. 
These personas will then be embedded into a specialized cognitive walkthrough (CW) protocol adapted from GenderMag \cite{wharton1994cognitive,burnett2016gendermag}. The CW forms will be adapted to explicitly incorporate the personas, with evaluation sheets and walkthrough questions restructured so that each decision point prompts evaluators to ask whether a persona (e.g., high masking, low tolerance for sensory load, strong need for structure) would be able to identify the correct action, interpret social cues, or manage the cognitive demands of the task. 

\textit{
\textbf{Step Three: Apply – Practitioner-Led Application of the Method with ND-Women Validation.}} In this phase, the method will be exercised with industry practitioners, including developers, team leaders, agile coaches, UX experts, and tool designers, who are not ND-Women. We will run remote workshops with industry teams to apply the adapted walkthrough on realistic tasks (e.g., sprint planning, code review, onboarding). Practitioners are targeted because they control day-to-day process changes (ceremonies, review norms, and evaluation). To ensure lived-experience fidelity, we will also run short, remote validation sessions with ND-women practitioners on the same scenarios and outputs. Validation may be synchronous or asynchronous (chat-only or written feedback), with accommodations for privacy and sensory needs (pseudonyms, cameras-off, breaks). ND-women feedback will confirm, refine, or reject issues raised by practitioner teams and can propose alternative adjustments. Outputs across both activities are a structured list of inclusiveness issues (task/tool, workflow, policy) and concrete changes to ceremonies, review templates, and collaboration norms.


\section{Where We Stand; Challenges Faced by Neurodivergent (Autistic \& ADHD) Women}

\setlength{\tabcolsep}{2.5pt}
\begin{table*}[h!]
\centering
\caption{Intersectional Challenges of Neurodivergent Women in Software Engineering.\label{tab:nd_women_challenges}}
\vspace*{-1mm}
{\footnotesize
\begin{tabular}{p{25mm}|p{35mm}|p{38mm}|p{35mm}|p{38mm}}
\toprule
\makecell[c]{\textbf{Categories/}\\\textbf{Domains}} &
\makecell[c]{\textbf{Core ND}\\\textbf{Challenges in SWE}} &
\makecell[c]{\textbf{Gendered}\\\textbf{Amplifiers (ND Women)}} &
\makecell[c]{\textbf{Impact}} &
\makecell[c]{\textbf{Examples}} \\
\midrule
\textbf{Cognitive / Executive} 
& Executive dysfunction: difficulty with planning, prioritisation, task estimation, working memory, inhibition \cite{Morris2015, Liebel2025, newman2025get, Grischa2024Challenges} 
& Later or missed diagnosis due to male-centric criteria; self-doubt, impostor feelings, lack of early accommodations \cite{hull2020female,LateDiagnosisBargielaEtAl2016, CamouflagingLaiEtAl2017, NDWomenHoldenEtAl2025} 
& Underdeveloped coping strategies; struggle to meet deadlines; chronic cognitive overload \cite{da2025felt, Costello2021Professional} 
& A ND women only diagnosed with ADHD in their 30s struggle to estimate sprint tasks accurately, leading managers to view them as “unreliable.” \\ \hline

\textbf{Social / Interpersonal} 
& Struggles with informal communication norms, eye contact, small talk; collaboration anxiety in agile teams \cite{Morris2015,NDInAgileTeamsGamaEtAl2023, Grischa2024Challenges} 
& Chronic masking/camouflaging to appear neurotypical; fear of stigma and disclosure, especially in male-dominated settings \cite {JessGoodman2023,Morris2015, Costello2021Professional, QualitativeStudyMilnerEtAl2019} 
& Social isolation, exclusion from peer learning networks, reduced visibility of contributions \cite{Costello2021Professional} 
& In pair programming, forcing eye contact and small talk drains energy, leaving less focus for coding tasks. \\ \hline

\textbf{Organizational / Environmental} 
& Distractions from open offices, excessive tools, multitasking; frequent interruptions disrupt focus \cite{Morris2015, Liebel2025, NDInAgileTeamsGamaEtAl2023, SasportesEtAl2024Challenges} 
& Lack of mentors and sponsors; stereotype threat leading to overcompensation and burnout \cite{da2025felt,JessGoodman2023, NeurodiversityHennekamEtAl2024, Hewlett2010sponsor} 
& Reduced productivity; difficulty sustaining performance; heightened stress and fatigue \cite{Grischa2024Challenges, da2025felt, Costello2021Professional} 
& Hot-desking and constant Slack pings overwhelm attention; ND women leave stand-ups exhausted before deep work begins. \\ \hline

\textbf{Structural / Cultural} 
& Lack of awareness or accommodations from managers; biased evaluation of performance \cite{Costello2021Professional, Morris2015, NeurodiversityHennekamEtAl2024}
& Intersectional barriers (gender $\times$ neurotype $\times$ race); underrepresentation in leadership roles \cite{JessGoodman2023,da2025felt} 
& Systemic exclusion; stagnated career progression; higher attrition risk \cite{da2025felt} 
& A SE developer missing daily stand-ups due to sensory overload is labelled “uncommitted,”. \\ \hline

\textbf{Career Pathway / Progression} 
& Inaccessible hiring practices (e.g., timed coding tests, rigid interviews, reliance on eye contact or small talk); biased promotion and performance review systems \cite{Costello2021Professional, Morris2015, SasportesEtAl2024Challenges} 
& Gendered expectations around confidence and assertiveness; lack of sponsorship opportunities; career breaks penalized more heavily for women \cite{Hewlett2010sponsor, Cohen2023gender, NeurodiversityHennekamEtAl2024} 
& Reduced access to leadership roles; stalled advancement; increased likelihood of leaving the profession \cite{da2025felt}
& An autistic SE professional woman performs strongly in technical problem-solving but is rejected after struggling with small talk in a panel interview. \\
\bottomrule
\end{tabular}
}
\vspace*{-2mm}
\end{table*}

We are currently undertaking the Scope phase of our broader research agenda, which serves as the foundation for our project. This stage involves a comprehensive literature review focused on the lived experiences of ND women in SE. Guided by two central questions:
\begin{enumerate}
\item \textbf{RQ1:} What are the main challenges faced by female ND software engineers in the industry?
\item \textbf{RQ2:} How can these challenges be categorized across cognitive, social, organizational, and structural domains?
\end{enumerate}

To address these RQs, we first identified the central challenges experienced by ND women in the SE industry \textbf{(RQ1)}. While many of these challenges are shared by neurodivergent practitioners more broadly, they are often intensified by gendered dynamics within a male-dominated industry (See Table~\ref{tab:nd_women_challenges}). To operationalize \textbf{RQ2}, we draw on prior work \cite{NDInAgileTeamsGamaEtAl2023, Grischa2024Challenges, SasportesEtAl2024Challenges, Morris2015} in neurodiversity studies and organizational research to define the domains as follows: \textit{Cognitive / Executive} refers to difficulties with planning, memory, and task management; \textit{Social / Interpersonal} encompasses struggles with informal communication norms and team dynamics; \textit{Organizational / Environmental} captures barriers in workplace setups such as open offices, multitasking, and lack of mentorship; \textit{Structural / Cultural} highlights systemic issues such as biased evaluations and insufficient accommodations; and \textit{Career Pathway / Progression} points to barriers in hiring, promotion, and access to leadership. These domains provide a structured lens for analyzing how challenges manifest and intersect across multiple levels of the software engineering workplace (See Table~\ref{tab:nd_women_challenges}). This structure emphasizes the main neurodivergent challenges and gendered amplifiers that contribute to exclusion, lower visibility, and potential attrition. Our analysis reveals a disconnect between typical SE processes and the experiences of ND women, indicating that mere 'add-on' accommodations fall short without changes to daily practices and evaluation systems.

\section{Reframing Software Engineering Through a Neurodiversity Lens}

The challenges outlined above do not occur in a vacuum, they intersect deeply with the fabric of modern software engineering practices, particularly in Agile-driven, collaborative environments \cite{NDInAgileTeamsGamaEtAl2023}. While designed for speed and efficiency, these practices embed implicit norms around communication and cognition that can disadvantage ND women. We propose that these frictions could prompt a rethinking of how SE is practiced, evaluated, and taught.

\textbf{Agile ceremonies and cadence:} Daily stand-ups and sprint rituals may increase cognitive load around planning, recall, and rapid turn-taking. When diagnosis comes later, women may enter teams without formal adjustments, so difficulties adapting to constant verbal updates can be read as “commitment” issues rather than signals to adjust process \cite{NDInAgileTeamsGamaEtAl2023,Morris2015,PresenttaionofAutismDriverEtAl2021,LateDiagnosisBargielaEtAl2016}. Apart from these, camouflaging to keep up could add fatigue \cite{hull2020female,CamouflagingLaiEtAl2017}.

\textbf{Code review and quality practices:} Modern code review can function as a participation and visibility gate. While many ND engineers excel at detail-oriented defect finding, unstructured expectations and implicit norms may penalise women’s contributions or tone and review work can also be high-load for ADHD (abstraction, working memory, time management) \cite{SasportesEtAl2024Challenges,newman2025get}. 

\textbf{Collaboration and culture:} Practices such as pair programming and continuous collaboration can be double-edged. They may support learning, but can also become exhausting when social interaction is constant. For ND women, the pressure to appear fully engaged and socially competent often leads to masking behaviour. This not only consumes additional energy but may also limit opportunities to build authentic relationships and access informal support networks \cite{NDInAgileTeamsGamaEtAl2023,da2025felt,QualitativeStudyMilnerEtAl2019,JessGoodman2023}. Over time, this cycle could reduce both well-being and visibility within the team.

These impacts suggest that exclusion does not come from lack of technical ability but from how current Agile practices can intensify late diagnosis, masking, and gaps in support for women. Adjustments in ceremonies, reviews, and collaboration could make participation more sustainable.
\section{Conclusion and Future Plans}
The challenges faced by ND women in SE are not edge cases rather they are signals of deeper structural issues in how the field defines communication, collaboration, and competence. Addressing these challenges demands more than accommodations bolted onto existing workflows. Our approach is drawing on InclusiveMag and GenderMag, is not just a toolkit but a methodological stance: one that prioritizes intersectionality, lived experience, and adaptability in analytic practice. In future work, we aim to develop \textit{NeurodiversiWMag} (Neurodivergent Women Inclusiveness Magnifier), a specialised methodological variant designed to center the experiences of neurodivergent women in SE. Our goal is to translate this framework into a reusable, practical resource for software teams, researchers, and educators. This will include validated persona libraries, adapted walkthrough prompts, and lightweight digital tools that help teams identify inclusion gaps without requiring formal facilitation. If SE is to serve all who build and use software, it must evolve beyond narrow norms of speed, social ease, and standardization.

\bibliographystyle{ACM-Reference-Format}
\bibliography{Main}

@inproceedings{NDInAgileTeamsGamaEtAl2023,
  title={Understanding and supporting neurodiverse software developers in agile teams},
  author={Gama, Kiev and Lacerda, Aline},
  booktitle={Proceedings of the XXXVII Brazilian Symposium on Software Engineering},
  pages={497--502},
  year={2023}
}

@article{NeurodiversityHennekamEtAl2024,
  title={Neurodiversity, gender, and work},
  author={Hennekam, Sophie and Hayward, Susan M and Bastian, Bettina Lynda},
  journal={Gender, Work \& Organization},
  year={2024}
}

@inproceedings{morreale2025we,
  title={How We Did It: Integrating Inclusive Design across the Undergraduate Computer Science Curriculum},
  author={Morreale, Patricia and Burnett, Margaret and Harms, Kyle J and Kwak, Daehan},
  booktitle={Proceedings of the 56th ACM Technical Symposium on Computer Science Education V. 2},
  pages={1703--1704},
  year={2025}
}

@article{hu2021toward,
  title={Toward a socioeconomic-aware HCI: Five facets},
  author={Hu, Catherine and Perdriau, Christopher and Mendez, Christopher and Gao, Caroline and Fallatah, Abrar and Burnett, Margaret},
  journal={arXiv preprint arXiv:2108.13477},
  year={2021}
}

@incollection{wharton1994cognitive,
  title={The cognitive walkthrough method: A practitioner's guide},
  author={Wharton, Cathleen and Rieman, John and Lewis, Clayton and Polson, Peter},
  booktitle={Usability inspection methods},
  pages={105--140},
  year={1994}
}

@article{jahandideh2025low,
  title={Low Battery Alarm; A Scoping Review of Autistic Burnout},
  author={Jahandideh, Pardis and Seyedmirzaei, Homa and Rasoulian, Pegah and Memari, Amirhossein},
  journal={Journal of Autism and Developmental Disorders},
  pages={1--21},
  year={2025},
  publisher={Springer}
}

@article{burnett2016gendermag,
  title={GenderMag: A method for evaluating software's gender inclusiveness},
  author={Burnett, Margaret and Stumpf, Simone and Macbeth, Jamie and Makri, Stephann and Beckwith, Laura and Kwan, Irwin and Peters, Anicia and Jernigan, William},
  journal={Interacting with computers},
  volume={28},
  number={6},
  pages={760--787},
  year={2016},
  publisher={Oxford University Press}
}

@inproceedings{letaw2021changing,
  title={Changing the online climate via the online students: Effects of three curricular interventions on online CS students’ inclusivity},
  author={Letaw, Lara and Garcia, Rosalinda and Garcia, Heather and Perdriau, Christopher and Burnett, Margaret},
  booktitle={Proceedings of the 17th ACM Conference on International Computing Education Research},
  pages={42--59},
  year={2021}
}

@article{rodriguez2021perceived,
  title={Perceived diversity in software engineering: a systematic literature review},
  author={Rodr{\'\i}guez-P{\'e}rez, Gema and Nadri, Reza and Nagappan, Meiyappan},
  journal={Empirical Software Engineering},
  volume={26},
  number={5},
  pages={102},
  year={2021},
  publisher={Springer}
}

@article{hoogman2020creativity,
  title={Creativity and ADHD: A review of behavioral studies, the effect of psychostimulants and neural underpinnings},
  author={Hoogman, Martine and Stolte, Marije and Baas, Matthijs and Kroesbergen, Evelyn},
  journal={Neuroscience \& Biobehavioral Reviews},
  volume={119},
  pages={66--85},
  year={2020},
  publisher={Elsevier}
}

@inproceedings{gama2023understanding,
  title={Understanding and supporting neurodiverse software developers in agile teams},
  author={Gama, Kiev and Lacerda, Aline},
  booktitle={Proceedings of the XXXVII Brazilian Symposium on Software Engineering},
  pages={497--502},
  year={2023}
}

@misc{Stack2022,
  author       = {{Stack Overflow}},
  title        = {2022 Stack Overflow Developer Survey},
  howpublished = {\url{https://survey.stackoverflow.co/2022/}},
  note         = {Accessed: 24 Sep. 2025},
  year         = {2022}
}

@article{austin2017neurodiversity,
  title={Neurodiversity as a competitive advantage},
  author={Austin, Robert D and Pisano, Gary P},
  journal={Harvard Business Review},
  volume={95},
  number={3},
  pages={96--103},
  year={2017}
}

@book{Grandin2022,
  author    = {Temple Grandin},
  title     = {Visual Thinking: The Hidden Gifts of People Who Think in Pictures, Patterns, and Abstractions},
  publisher = {Riverhead Books},
  year      = {2022},
}

@article{hennekam2024neurodiversity,
  title={Neurodiversity, gender, and work},
  author={Hennekam, Sophie and Hayward, Susan M and Bastian, Bettina Lynda},
  journal={Gender, Work \& Organization},
  year={2024}
}

@article{marquez2024inclusion,
  title={Inclusion of individuals with autism spectrum disorder in software engineering},
  author={M{\'a}rquez, Gast{\'o}n and Pacheco, Michelle and Astudillo, Hern{\'a}n and Taramasco, Carla and Calvo, Esteban},
  journal={Information and Software Technology},
  volume={170},
  pages={107434},
  year={2024},
  publisher={Elsevier}
}

@article{raymaker2020having,
  title={“Having all of your internal resources exhausted beyond measure and being left with no clean-up crew”: Defining autistic burnout},
  author={Raymaker, Dora M and Teo, Alan R and Steckler, Nicole A and Lentz, Brandy and Scharer, Mirah and Delos Santos, Austin and Kapp, Steven K and Hunter, Morrigan and Joyce, Andee and Nicolaidis, Christina},
  journal={Autism in adulthood},
  volume={2},
  number={2},
  pages={132--143},
  year={2020},
  publisher={Mary Ann Liebert, Inc., publishers 140 Huguenot Street, 3rd Floor New~…}
}

@article{mantzalas2022autistic,
  title={What is autistic burnout? A thematic analysis of posts on two online platforms},
  author={Mantzalas, Jane and Richdale, Amanda L and Adikari, Achini and Lowe, Jennifer and Dissanayake, Cheryl},
  journal={Autism in Adulthood},
  volume={4},
  number={1},
  pages={52--65},
  year={2022},
  publisher={Mary Ann Liebert, Inc., publishers 140 Huguenot Street, 3rd Floor New~…}
}

@inproceedings{mcintosh2021evaluating,
  title={Evaluating age bias in e-commerce},
  author={McIntosh, Jennifer and Du, Xiaojiao and Wu, Zexian and Truong, Giahuy and Ly, Quang and How, Richard and Viswanathan, Sriram and Kanij, Tanjila},
  booktitle={2021 IEEE/ACM 13th International Workshop on Cooperative and Human Aspects of Software Engineering (CHASE)},
  pages={31--40},
  year={2021},
  organization={IEEE}
}

@inproceedings{mendez2019gendermag,
  title={From GenderMag to InclusiveMag: An inclusive design meta-method},
  author={Mendez, Christopher and Letaw, Lara and Burnett, Margaret and Stumpf, Simone and Sarma, Anita and Hilderbrand, Claudia},
  booktitle={2019 IEEE Symposium on Visual Languages and Human-Centric Computing (VL/HCC)},
  pages={97--106},
  year={2019},
  organization={IEEE}
}

@misc{AUsupport,
  year={2024},
url={https://dxc.com/au/en/newsroom/07152024?utm_source=chatgpt.com}, journal={DXC Technology} }

@article{dwyer2022neurodiversity,
  title={The neurodiversity approach (es): What are they and what do they mean for researchers?},
  author={Dwyer, Patrick},
  journal={Human development},
  volume={66},
  number={2},
  pages={73--92},
  year={2022}
}

@article{QualitativeStudyMilnerEtAl2019,
  title={A qualitative exploration of the female experience of autism spectrum disorder (ASD)},
  author={Milner, Victoria and McIntosh, Hollie and Colvert, Emma and Happ{\'e}, Francesca},
  journal={Journal of autism and developmental disorders},
  volume={49},
  number={6},
  pages={2389--2402},
  year={2019},
  publisher={Springer}
}

@article{LateDiagnosisBargielaEtAl2016,
  title={The experiences of late-diagnosed women with autism spectrum conditions: An investigation of the female autism phenotype},
  author={Bargiela, Sarah and Steward, Robyn and Mandy, William},
  journal={Journal of autism and developmental disorders},
  volume={46},
  number={10},
  pages={3281--3294},
  year={2016},
  publisher={Springer}
}

@article{NDWomenHoldenEtAl2025,
  title={Adverse experiences of women with undiagnosed ADHD and the invaluable role of diagnosis},
  author={Holden, Eve and Kobayashi-Wood, Helena},
  journal={Scientific Reports},
  volume={15},
  number={1},
  pages={20945},
  year={2025},
  publisher={Nature Publishing Group UK London}
}

@article{IsueswithFemalescCiliaEtAl2023,
  title={Issues which marginalize females with ADHD-A mixed methods systematic review},
  author={Cilia Vincenti, Sarah and Galea, Michael and Briffa, Vince},
  year={2023},
  publisher={IDEAS SPREAD}
}

@article{PresenttaionofAutismDriverEtAl2021,
  title={The presentation, recognition and diagnosis of autism in women and girls},
  author={Driver, Bethany and Chester, Verity},
  journal={Advances in Autism},
  volume={7},
  number={3},
  pages={194--207},
  year={2021},
  publisher={Emerald Publishing Limited}
}

@article{AustisticWomenChesterEtAl2019,
  title={Autistic women and girls: Increasingly recognised, researched and served},
  author={Chester, Verity},
  journal={Advances in Autism},
  volume={5},
  number={3},
  pages={141--142},
  year={2019},
  publisher={Emerald Publishing Limited}
}

@article{SexDifferencesPengEtAl2023,
  title={Sex differences in the manifestation of ADHD: A longitudinal examination},
  author={Peng, Zheyue and Watts, Ashley L},
  year={2023},
  publisher={OSF}
}

@article{CamouflagingLaiEtAl2017,
  title={Quantifying and exploring camouflaging in men and women with autism},
  author={Lai, Meng-Chuan and Lombardo, Michael V and Ruigrok, Amber NV and Chakrabarti, Bhismadev and Auyeung, Bonnie and Szatmari, Peter and Happ{\'e}, Francesca and Baron-Cohen, Simon and MRC AIMS Consortium},
  journal={Autism},
  volume={21},
  number={6},
  pages={690--702},
  year={2017},
  publisher={Sage Publications Sage UK: London, England}
}

@article{JessGoodman2023,
  title={Neurodivergence and Marginalised Gender-a thematic analysis of womens’ and gender-diverse peoples’ experiences of ASD and ADHD.},
  author={Goodman, Jess},
  year={2025},
  publisher={OSF}
}

@article{Liebel2025,
  title={Differences between Neurodivergent and Neurotypical Software Engineers: Analyzing the 2022 Stack Overflow Survey},
  author={Verma, Pragya and Cruz, Marcos Vinicius and Liebel, Grischa},
  journal={arXiv preprint arXiv:2506.03840},
  year={2025}
}

@inproceedings{Morris2015,
  title={Understanding the challenges faced by neurodiverse software engineering employees: Towards a more inclusive and productive technical workforce},
  author={Morris, Meredith Ringel and Begel, Andrew and Wiedermann, Ben},
  booktitle={Proceedings of the 17th International ACM SIGACCESS Conference on computers \& accessibility},
  pages={173--184},
  year={2015}
}

@article{SasportesEtAl2024Challenges,
  title={Challenges and Opportunities for Neurodivergent Software Engineers Modern Code Reviews and Bug Finding},
  author={Sasportes, Madalena Ribas},
  journal={PQDT-Global},
  year={2024},
  publisher={Universidade NOVA de Lisboa (Portugal)}
}

@inproceedings{Costello2021Professional,
  title={A professional career with autism: Findings from a literature review in the software engineering domain},
  author={Costello, Emma and Kilbride, Sara and Milne, Zoe and Clarke, Paul and Yilmaz, Murat and MacMahon, Silvana Togneri},
  booktitle={European Conference on Software Process Improvement},
  pages={349--360},
  year={2021},
  organization={Springer}
}

@book{Hewlett2010sponsor,
  title={The sponsor effect: Breaking through the last glass ceiling},
  author={Hewlett, Sylvia Ann and Peraino, Kerrie and Sherbin, Laura and Sumberg, Karen and others},
  year={2010},
  publisher={Harvard Business Review Boston, MA}
}

@book{Cohen2023gender,
  title={The Gender Bias: The Barriers that Hold Women Back, and how to Break Them},
  author={Cohen-Hatton, Sabrina},
  year={2023},
  publisher={Kings Road Publishing}
}

@inproceedings{Grischa2024Challenges,
author = {Liebel, Grischa and Langlois, Noah and Gama, Kiev},
title = {Challenges, Strengths, and Strategies of Software Engineers with ADHD: A Case Study},
year = {2024},
isbn = {9798400704994},
publisher = {Association for Computing Machinery},

doi = {10.1145/3639475.3640107},
booktitle = {Proceedings of the 46th International Conference on Software Engineering: Software Engineering in Society},
pages = {57–68},
numpages = {12},
series = {ICSE-SEIS'24}
}

@article{kapp2013deficit,
  title={Deficit, difference, or both? Autism and neurodiversity.},
  author={Kapp, Steven K and Gillespie-Lynch, Kristen and Sherman, Lauren E and Hutman, Ted},
  journal={Developmental psychology},
  volume={49},
  number={1},
  pages={59},
  year={2013},
  publisher={American Psychological Association}
}

@misc{den2019neurodiversity,
  title={Neurodiversity: An insider’s perspective},
  author={Den Houting, Jacquiline},
  journal={Autism},
  volume={23},
  number={2},
  pages={271--273},
  year={2019},
  publisher={Sage Publications Sage UK: London, England}
}

@book{armstrong2010neurodiversity,
  title={Neurodiversity: Discovering the extraordinary gifts of autism, ADHD, dyslexia, and other brain differences},
  author={Armstrong, Thomas},
  year={2010},
  publisher={ReadHowYouWant. com}
}

@article{hull2020female,
  title={The female autism phenotype and camouflaging: A narrative review},
  author={Hull, Laura and Petrides, KV and Mandy, William},
  journal={Review journal of autism and developmental disorders},
  volume={7},
  number={4},
  pages={306--317},
  year={2020},
  publisher={Springer}
}

@article{cook2021camouflaging,
  title={Camouflaging in autism: A systematic review},
  author={Cook, Julia and Hull, Laura and Crane, Laura and Mandy, William},
  journal={Clinical psychology review},
  volume={89},
  pages={102080},
  year={2021},
  publisher={Elsevier}
}

@misc{pohl2020comparative,
  title={A comparative study of autistic and non-autistic women’s experience of motherhood. Molecular Autism, 11 (3)},
  author={Pohl, AL and Crockford, SK and Blakemore, M and Allison, C and Baron-Cohen, S},
  year={2020}
}

@inproceedings{newman2025get,
  title={“Get Me In The Groove”: A Mixed Methods Study on Supporting ADHD Professional Programmers},
  author={Newman, Kaia and Snay, Sarah and Endres, Madeline and Parikh, Manasvi and Begel, Andrew},
  booktitle={2025 IEEE/ACM 47th International Conference on Software Engineering (ICSE)},
  pages={778--778},
  year={2025},
  organization={IEEE Computer Society}
}

@inproceedings{da2025felt,
  title={“I Felt Pressured to Give 100\% All the Time”: How Are Neurodivergent Professionals Being Included in Software Development Teams?},
  author={da Silva Menezes, Nicoly and da Rocha, Thayssa {\'A}guila and Camelo, Lucas Samuel Santiago and Mota, Marcelle Pereira},
  booktitle={Simp{\'o}sio Brasileiro de Sistemas de Informa{\c{c}}{\~a}o (SBSI)},
  pages={525--534},
  year={2025},
  organization={SBC}
}

@article{graafsma2022no,
  title={No evidence that autistic traits predict programming learning outcomes},
  author={Graafsma, Irene L and Marinus, Eva and Robidoux, Serje and Nickels, Lyndsey and Caruana, Nathan},
  journal={Computers in Human Behavior Reports},
  volume={7},
  pages={100215},
  year={2022},
  publisher={Elsevier}
}

\appendix

\end{document}